\documentclass[aps,aps, llncs, twocolumn]{revtex4}
\usepackage{amsmath}
\usepackage{amsfonts}
\usepackage{amssymb}
\usepackage{graphicx}
\usepackage{amsmath}
\usepackage{float}
\usepackage{color}
\usepackage{soul}
\usepackage{siunitx}
\usepackage{physics}
\usepackage[utf8]{inputenc}
\usepackage[T1]{fontenc}

\DeclareUnicodeCharacter{0096}{_}

\begin{document}
\title{Spin-polarized supercurrent through the van der Waals Kondo lattice ferromagnet Fe$_3$GeTe$_2$} % Article title

\author{Deepti Rana$^1$, Aswini R$^1$, Basavaraja G$^2$, Chandan Patra$^3$,  Sandeep Howlader$^1$, Rajeswari Roy Chowdhury$^3$, Mukul Kabir$^2$, Ravi P. Singh$^3$, and Goutam Sheet$^1$} 

\email{goutam@iisermohali.ac.in}

\affiliation{$^1$Department of Physical Sciences, Indian Institute of Science Education and Research (IISER) Mohali, Sector 81, S. A. S. Nagar, Manauli, PO 140306, India}

\affiliation{$^2$Department of Physics, Indian Institute of Science Education and Research (IISER) Pune, PO 411008, India}

\affiliation{$^3$Department of Physics, Indian Institute of Science Education and Research (IISER) Bhopal, PO 462066, India}

\begin{abstract}
\textbf{In the new van der Waals Kondo-lattice Fe$_3$GeTe$_2$, itinerant ferromagnetism and heavy fermionic behaviour coexist. Both the key properties of such a system namely a spin-polarized Fermi surface and a low Fermi momentum are expected to significantly alter Andreev reflection dominated transport at a contact with a superconducting electrode, and display unconventional proximity-induced superconductivity. We observed interplay between Andreev reflection and Kondo resonance at mesoscopic interfaces between superconducting Nb and Fe$_3$GeTe$_2$. Above the critical temperature ($T_c$) of Nb, the recorded differential conductance ($dI/dV$) spectra display a robust zero-bias anomaly which is described well by a characteristic Fano line shape arising from Kondo resonance. Below $T_c$, the Fano line mixes with Andreev reflection dominated $dI/dV$ leading to a dramatic, unconventional suppression of conductance at zero bias. As a consequence, an analysis of the Andreev reflection spectra within a spin-polarized model yields an anomalously large spin-polarization which is not explained by the density of states of the spin-split bands at the Fermi surface alone. The results open up the possibilities of fascinating interplay between various quantum phenomena that may potentially emerge at the mesoscopic superconducting interfaces involving Kondo lattice systems hosting spin-polarized conduction electrons. }
\end{abstract}

\maketitle

Spin-polarized transport characteristics through mesoscopic junctions between conventional superconductors and complex magnetic systems may display unconventional features related to exotic Fermi surface properties. In a normal metal/superconducting junction the transport is dominated by Andreev reflection \cite{Andreev}, a quantum process through which conversion of normal-current in the metal to a supercurrent in the superconductor happens. The process involves reflection of an up (down) spin electron as a down (up) spin hole thereby causing a conductance enhancement below the superconducting energy gap. When the normal metal is an itinerant ferromagnet characterized by a spin-polarized Fermi surface, all the electrons in the Fermi surface cannot undergo Andreev reflection  as all the corresponding Andreev reflected holes would not find states in the opposite spin band. This leads to a suppression of Andreev reflection in spin-polarized junctions. A measurement of the suppression of Andreev reflection gives an estimate of the spin polarization at the Fermi surface \cite{Soulen,Jong,Upadhyay,Auth,Howlader,Kamboj,Aggarwal,Sirohi,Mukhopadhyay}. The problem gains additional complexity when the spin-polarized electrode forming the junction with a superconductor hosts significant electron correlations. In such cases, additional suppression of Andreev reflection is expected due to a larger effective potential barrier that would arise from the intrinsic mismatch in the Fermi velocities in the two sides of the junction \cite{Tuuli}. The situation approaches an extreme limit when the carriers in the spin-polarized part of the junction are also characterized by a large effective mass as in the heavy Fermions. In such cases, the suppression of conductance may not be only due to spin polarization and larger potential barrier, both of which can be modelled within an appropriately modified version of the conventional Blonder-Tinkham-Klapwijk (BTK) theory\cite{Soulen}. 

The transport spectroscopic features in the afore-mentioned situations may also involve signature of other electronic effects in various forms, including a Kondo anomaly \cite{Park,Fogelstrom,Park3,Yang2}. Within the theory of electron tunneling into a Kondo lattice\cite{Maltseva} (as in case of a point contact geometry), it was earlier shown that a co-tunneling mechanism causes spin-flip processes. In presence of that, the calculation of the conductance within a mean field picture predicts the appearance of two peaks separated by a hybridization gap in the clean limit which gets smeared out as one approaches the dirty (disordered) limit. Within this picture, even with moderate disorder, the conductance spectrum is expected to take the shape of a Fano line \cite{Fano}. Such Fano lineshape behaviour were experimentally observed in point contacts with a number of heavy fermion systems in the past \cite{Park,Park3, Fogelstrom,Yang2}. 

In addition to the above, other complex possibilities may also arise. For example, the superconducting phase induced (proximity) in the heavy fermionic part of the junction may achieve unconventional character in the order parameter symmetry. While such possibilities were explored experimentally in non-magnetic or weakly magnetic heavy fermion superconductors like CeCoIn$_5$ \cite{Park}, CeCu$_2$Si$_2$, URu$_2$Si$_2$\cite{Wilde}, UBe$_{13}$, UPt$_3$ \cite{Nowack}, UTe$_2$ \cite{Jiao}, etc., investigation of such phenomena in a ferromagnetic Kondo lattice with heavy fermionic character was not investigated, mainly due to lack of a model system where all such physical properties would coexist. 

Recently, it was shown that the van der Waals ferromagnet Fe$_3$GeTe$_2$ hosts Fermi surface spin-polarization, an emergent Kondo lattice behaviour, along with a large carrier mass leading to a heavy fermion character to the system \cite{Zhang,Zhao}. Bulk crystalline Fe$_3$GeTe$_2$ is a van der Waals layered material that crystallizes into a hexagonal lattice structure with spacegroup $P6_3$/mmc \cite{Deiseroth,Chen}, as shown in Figure 1(a). The Fe atoms occupy two inequivalent Wyckoff positions denoted as Fe-I and Fe-II. The Fe-II atoms are covalently bonded with Ge at the middle layer, which is sandwiched between two hexagonal layers of Fe-I. This triple-layer Fe$_3$Ge are further sandwiched between two hexagonal Te layers, and a van der Waals gap separates the resultant pentuple layers Fe$_3$GeTe$_2$. Unlike other bulk vdW ferromagnets CrI$_3$ \cite{Huang}and CrSiTe$_3$ \cite{Lin}, it is an itinerant ferromagnet with high Curie temperature of 220 K-230 K \cite{Deiseroth,Chen} which can be further increased by doping \cite{Deng} or patterning \cite{Li}, making it a promising candidate for next-generation spintronic devices \cite{Burch,Mak,Gong}. The system shows planar topological Hall effect \cite{You}, along with significantly high uniaxial magnetic anisotropy \cite{Leon-Brito,Zhuang,Verchenko,Chen2}. Existence of strong correlations in Fe$_3$GeTe$_2$ are concluded from evidences of enhanced specific heat described by a high Sommerfeld coefficient pointing to a ($\sim$ 10 folds) mass enhancement \cite{Zhu}. In this work, we have performed Andreev reflection spectroscopy at point-contact junctions between tips of superconducting Nb and single crystals of the Kondo lattice ferromagnet Fe$_3$GeTe$_2$.

High quality single crystals of Fe$_3$GeTe$_2$ were synthesized by chemical vapor transport. The details of the growth parameters and the characterization are reported elsewhere \cite{Chowdhury}. Owing to the van der Waals bonding between the different layers, the single crystals were cleavable by mechanical exfoliation. We exposed fresh surface of the crystals before transferring them to the cryogenic measurement stages. 

\begin{figure}[h!]
 \centering
  \includegraphics[scale=0.17]{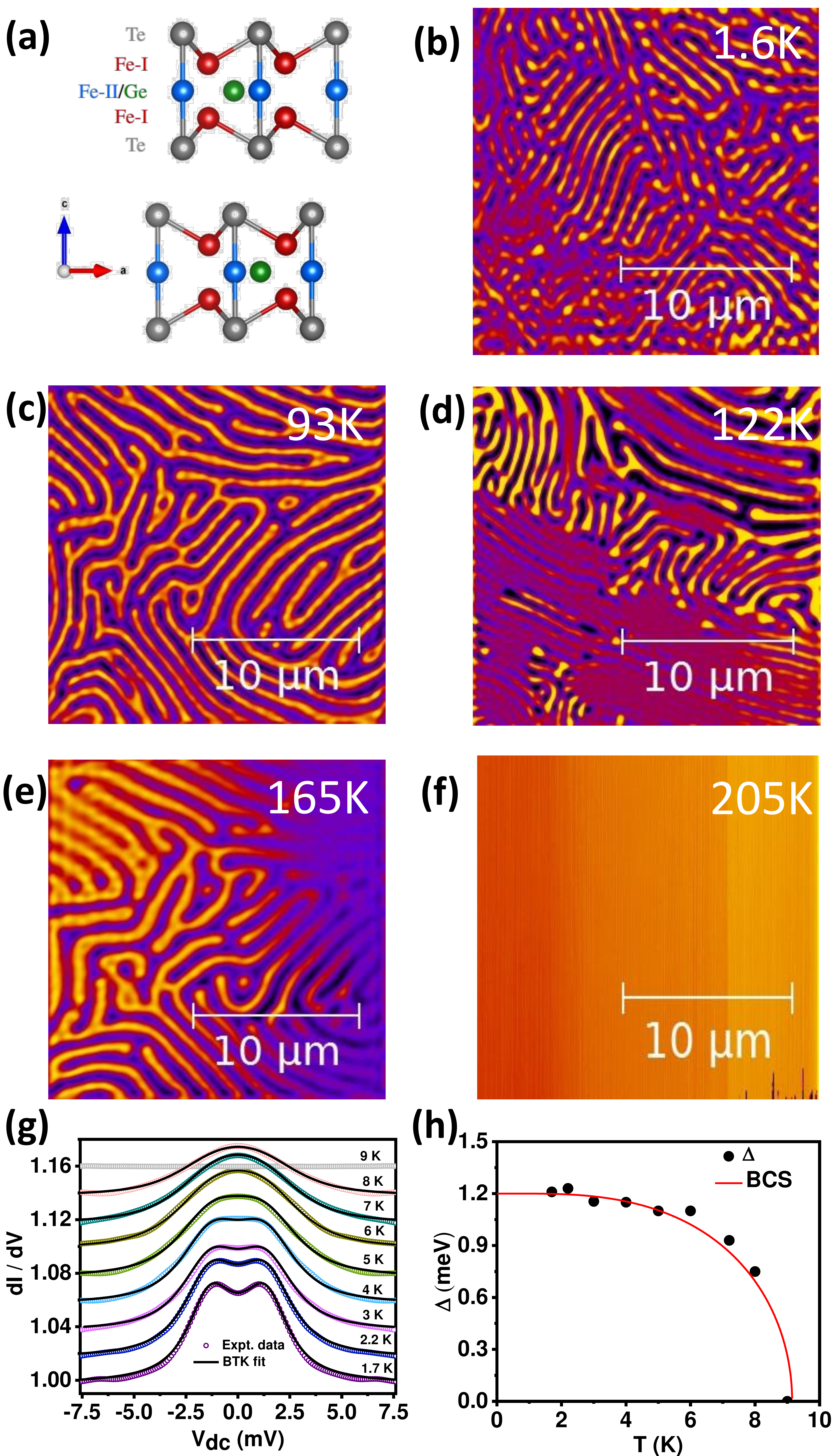}

	\caption{{(a) Crystal structure of Fe$_3$GeTe$_2$. (b)-(f) MFM dual pass phase images of Fe$_3$GeTe$_2$, at different temperatures, showing  contrast of ferromagnetic domains. The scale bar is 10$\mu$m. The stripe domains vanishes as temperature is raised to 205K. (g) Temperature dependence of ballistic spectra (shown by colored dots) and their  corresponding BTK fits (shown by black line). All the spectra are normalised and equal vertical shift to spectra with respect to conductance spectrum at 1.7K is given for clarity. (h) Temperature dependence of the superconducting gap (shown by black dots). The expected variation of the gap from  BCS theory is shown by solid red line.   }}
\end{figure}

Magnetization measurements revealed a critical temperature of $\sim$ 206 K \cite{Chowdhury}. In order to investigate the local magnetic properties, we imaged the ferromagnetic domains of Fe$_3$GeTe$_2$ directly by low-temperature magnetic force microscopy. At the lowest temperature, in absence of any external magnetic field treatment, we found clear stripe domains with a typical stripe width of $\sim$ 700 nm (Figure 1(b)). The stripe domains in a large size crystal is a consequence of strong uniaxial anisotropy, as is usually reported for Fe$_3$GeTe$_2$ from bulk magnetization measurements \cite{Leon-Brito,Cai,Wang}. The stripe domains also form interconnects closely resembling the domain structures that are theoretically expected for the zero-field treated states of skyrmionic systems. We have also performed field-cooled measurements to investigate the skyrmion physics in the system, but a discussion of that is beyond the scope of this paper. The temperature evolution of the ferromagnetic domains is presented in Figure 1(b)-(f). The width of the stripe domains increases with increase in temperature and disappears at 205 K, near the Curie temperature. The temperature range over which Andreev reflection experiments discussed below were performed, no significant change in the domain structure/size was seen. As we discuss below, statistically, the large domain size enabled the Nb tips to engage on individual domains for spin-polarization measurements through Andreev reflection.

\begin{figure}[h!]
 \centering
  \includegraphics[scale=0.18]{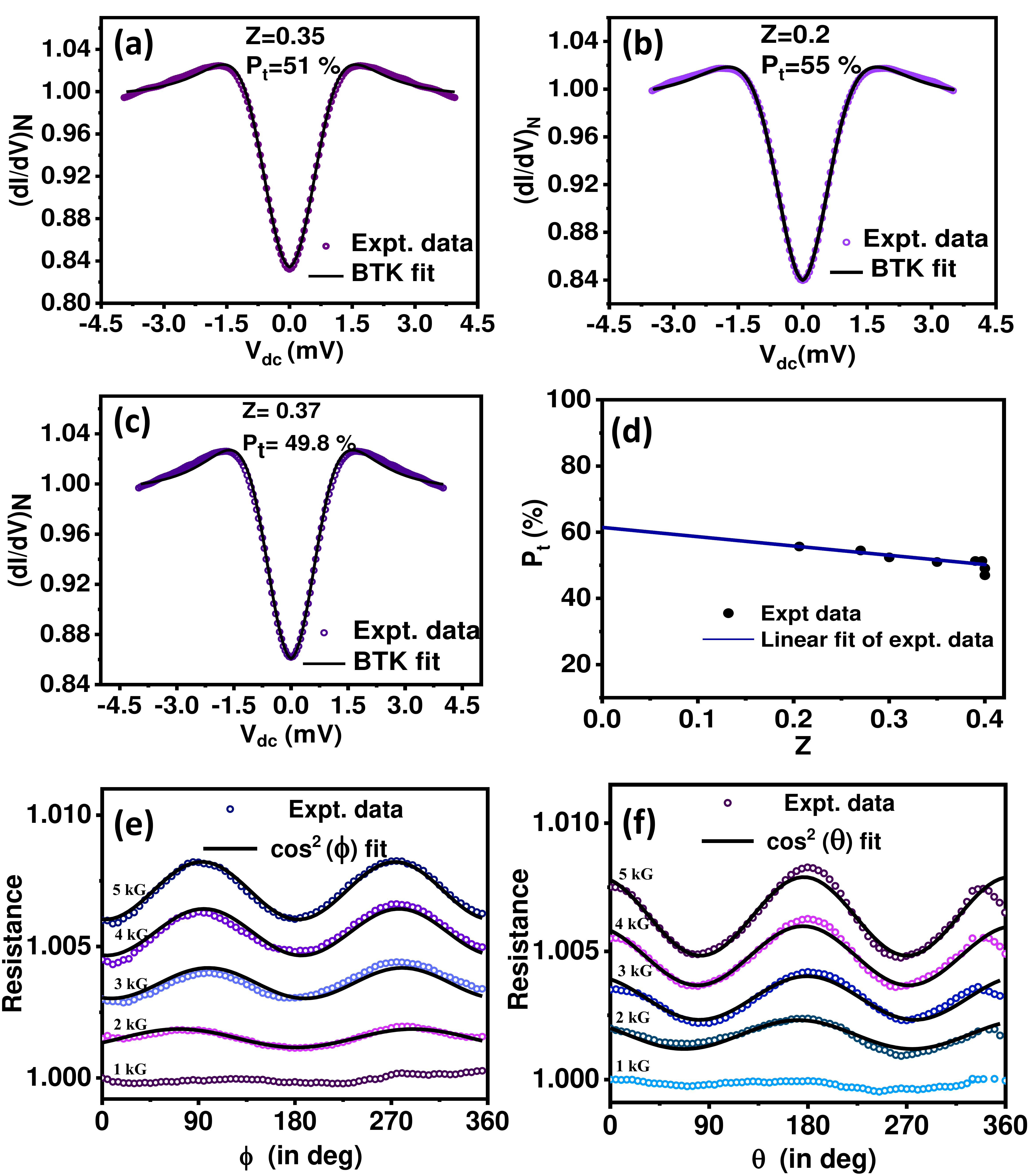}

	\caption{{(a)-(c) Three fitted ballistic spectra with different values of barrier strength (Z) and spin polarisation ($P_t$). (d) $P_t$ vs Z. The extrapolated value of $P_t$ (at Z=0) is around 61\%. (e)-(f) In-plane and out-of plane magnetic field angle dependence of resistance taken at zero bias and 11K. The $\cos^2 (\phi)$ and $\cos^2 (\theta)$ fits are shown by solid black lines. All the resistance curves are normalised and equal vertical shift to resistance curves with respect to curve at lowest magnetic field is given for clarity.
  }}
\end{figure}

For the Andreev reflection spectroscopy experiments, a Nb tip was engaged by standard needle-anvil method on a freshly cleaved single crystal of Fe$_3$GeTe$_2$ inside a variable temperature cryostat working down to 1.4 K which is also equipped with a 3-axis superconducting vector magnet (6T-1T-1T). The point contacts were formed on the [001] facet such that the current was injected along the c-axis of the crystal. For this direction of current injection, the layered structure does not play a role in deciding
the point contact resistance and its microscopic anatomy Owing to the high quality of the single crystals, ballistic superconducting point contacts, characterized by two differential conductance ($dI/dV$) peaks symmetric about $V$ = 0, could be established fairly easily. One such representative spectrum and its temperature dependence is shown in Figure 1(g). The colored points are experimentally obtained data points at different temperatures and the solid black lines are the corresponding fits within a modified BTK theory \cite{Soulen}-- modified to incorporate the effect of the spin-polarized band structure of the ferromagnetic fraction of the point contacts. 

In a normal metal-superconductor Andreev reflection process, the zero-bias conductance should be 2 times the normal state conductance, for a fully transparent barrier and at absolute zero \cite{Blonder}. While finite temperature broadens the spectral features with marginal reduction of $dI/dV$ at $T<0.5T_c$, a non-zero interfacial barrier ($Z\neq$0) causes suppression of the zero-bias conductance in a characteristic way that also causes enhancements (sharpening) of the $dI/dV$ peaks near $V = \pm \Delta/e$. However, a visual inspection of the spectra presented in Figure 1(g) it is clear that the barrier is transparent (low $Z$), but despite that Andreev reflection has been suppressed significantly. The zero-bias enhancement is only about 6\%. The spectrum could be described well within the modified BTK theory with an effective spin polarization ($P_t$) of 46.57\%, but only with an enhanced effective temperature of 4.6 K, or a rather large (Dynes like \cite{Dynes}) broadening parameter ($\Gamma$) approaching 0.52 meV which is almost 0.5$\Delta$. The need of a significantly enhanced effective temperature, or, instead, a larger $\Gamma$, is due to additional broadening effects that could be playing a role here, but the origin of which is unclear. We note that this effect is not due to contact heating because the normal state resistance did not change with increasing temperature, no additional spectral features other than the double-peak structure was seen, and Wexler's formula \cite{Wexler} gave an estimate of the contact diameter $\sim$ 20 nm which is smaller than the mean free path in Fe$_3$GeTe$_2$ thereby confirming that the contacts are ballistic and no significant contact heating is expected. Since the contact diameter is smaller than the domain size, statistically, majority of the times, the point contacts are formed on single domains. Furthermore, taking the measured temperature as the contact temperature, the $\Delta$ vs. $T$ graph is well described by the Bardeen-Cooper-Schrieffer (BCS) theory \cite{Bardeen} (solid red line in Figure 1(h)). It should be noted that a significant $\Delta$ is found even at measured $T$ = 8K -- any significant contact heating would make the contact non-superconducting at a much lower temperature. Furthermore, a larger effective temperature should also lead to an underestimation of $P_t$. There is also a possibility that certain local disorders under the point contacts, giving rise to the additional broadening. The point contacts were made on the freshly cleaved surfaces of single crystalline Fe$_3$GeTe$_2$. Hence, the possibility of such a effect should be low -- though that cannot be completely ruled out.

As per the standard practice, the intrinsic spin-polarization can be estimated by performing experiments with a number of spectra for junctions with different barrier strength ($Z$), and then extrapolating the $Z$-dependence of $P_t$ to $Z$ = 0. We investigated several other point contacts which display features of a higher $Z$. We show three such representative spectra, along with their modified BTK fits in Figure 2 (a-c). $P_t$ was seen to monotonically decrease with increasing $Z$ and the extrapolated dependence to $Z$ = 0 revealed a spin polarization greater than 60\% as shown in Figure 2(d). In all these cases, however, the effective temperature (or, the artificially introduced Dyne's-like broadening parameter $\Gamma$) used for the analysis was significantly high, indicating that the intrinsic Fermi surface spin polarization could be even higher.  

Here we would also like to highlight that the point contact showed strong anisotropy in magnetoresistance. To investigate that, we performed field-angle dependence of the normal state resistance of one such ballistic point contact at zero bias. The orientation of the  magnetic field with respect to the applied current was varied using the 3-axis vector magnet. The results for in-plane rotation ($\phi$) of field and out-of-plane rotation ($\theta$) of the magnetic field are shown in Figure 2(e) and (f) respectively. The anisotropy in magnetoresistance became more pronounced with increase in the strength of the field. The anisotropic behavior followed typical $cos^2(\phi)$ and $cos^2(\theta)$ dependence (shown by the black lines in the Figure 2(e)-(f)) signifying spin-polarized nature \cite{Aggarwal} of the transport super current flowing through the Nb/Fe$_3$GeTe$_2$ point contacts.

In order to gain insight on the spin-polarized Fermi surface of Fe$_3$GeTe$_2$, we performed first-principles density functional calculations. We have presented the detailed calculated band structures in Figure 3(a). We focus on the key features of the calculated band structure here. The spin-polarized electronic structure analysis indicates the metallic nature of Fe$_3$GeTe$_2$, which along with the non-integer magnetic moment, supports the itinerant nature of ferromagnetism. The appearance of several flat bands near the Fermi level indicates enhanced quasiparticle mass suggesting a strong electron correlation in the material. This is in agreement with the high Sommerfeld coefficient of specific heat data published in the past \cite{Zhu}. The states at the Fermi level are constructed with $d_{yz}$ and $d_{xz}$ orbitals from Fe-I hybridized with the Te-$p$ orbitals for the majority spin channel. In addition, a contribution from the $d_{xy}$ from Fe-II is also observed but is significantly lower. Similar trends are encountered for the minority spin channel. To note, the band structure of the system was also calculated earlier \cite{Zhang}, though not in the context of the Fermi level spin-polarization. The key aspects of our calculations are consistent with the past calculations. We have used the calculated bands to extract additional parameters for the analysis of our experimental data.
In general, as far as the transport measurements are concerned, the general expression for (transport) spin polarization can be written as \cite{Mazin}
\begin{equation}
P_t^n = \frac{\langle N(E_{\rm F})v^n_{\rm F}\rangle_{\uparrow} - \langle N(E_{\rm F})v^n_{\rm F}\rangle_{\downarrow}}{\langle N(E_{\rm F})v^n_{\rm F}\rangle_{\uparrow} + \langle N(E_{\rm F})v^n_{\rm F}\rangle_{\downarrow}},   
\end{equation}  
where $v_{\rm F}$ is the spin-polarised Fermi velocity. The Fermi velocity of the individual bands can be calculated from the slope of the individual bands at the Fermi energy. Taking the arithmetic average of the velocities of all the bands results in the the average Fermi velocity for the respective spin channels. $n=0$ gives the net Fermi surface spin polarization which is not the relevant quantity here, as the Fermi velocity of the up and the down spin channels can be different too. In point-contact spectroscopy $n=1$ and $n=2$ give $P_t$ in ballistic and diffusive regimes respectively. From our calculations, we found the average Fermi velocity of the majority spin channel to be $v_{{\rm F}\uparrow}=\SI{4.10e5} {\meter\per\second}$, which is larger than the minority spin channel, $v_{{\rm F}\downarrow}=\SI{2.86e5} {\meter\per\second}$. The observation of Fermi velocities an order of magnitude slower than that in the typical metals is consistent with the reported higher Fermion mass in the system. The calculated spin-polarized density of states at the Fermi level is $N_{\uparrow}=\SI{4.72}{states/\electronvolt/unitcell}$, that is $79\%$ greater than $N_{\downarrow}$ [Figure3(b)]. Therefore, the imbalance of both $v_{\rm F}$ and $N(E_{\rm F})$ result into the transport spin polarization of $44\%$ in the ballistic regime and $58\%$ in the diffusive regime. To note, these estimates were done at absolute zero temperature and did not involve the effects of thermal broadening at the measurement temperatures. In principle, the measured spin-polarization should be significantly smaller than the theoretically obtained numbers. However, the experimentally measured spin polarization was found to be higher even with larger effective temperatures (or higher effective broadening parameters ($\Gamma$)).
\begin{figure}[h!]
 \centering
  \includegraphics[scale=0.098]{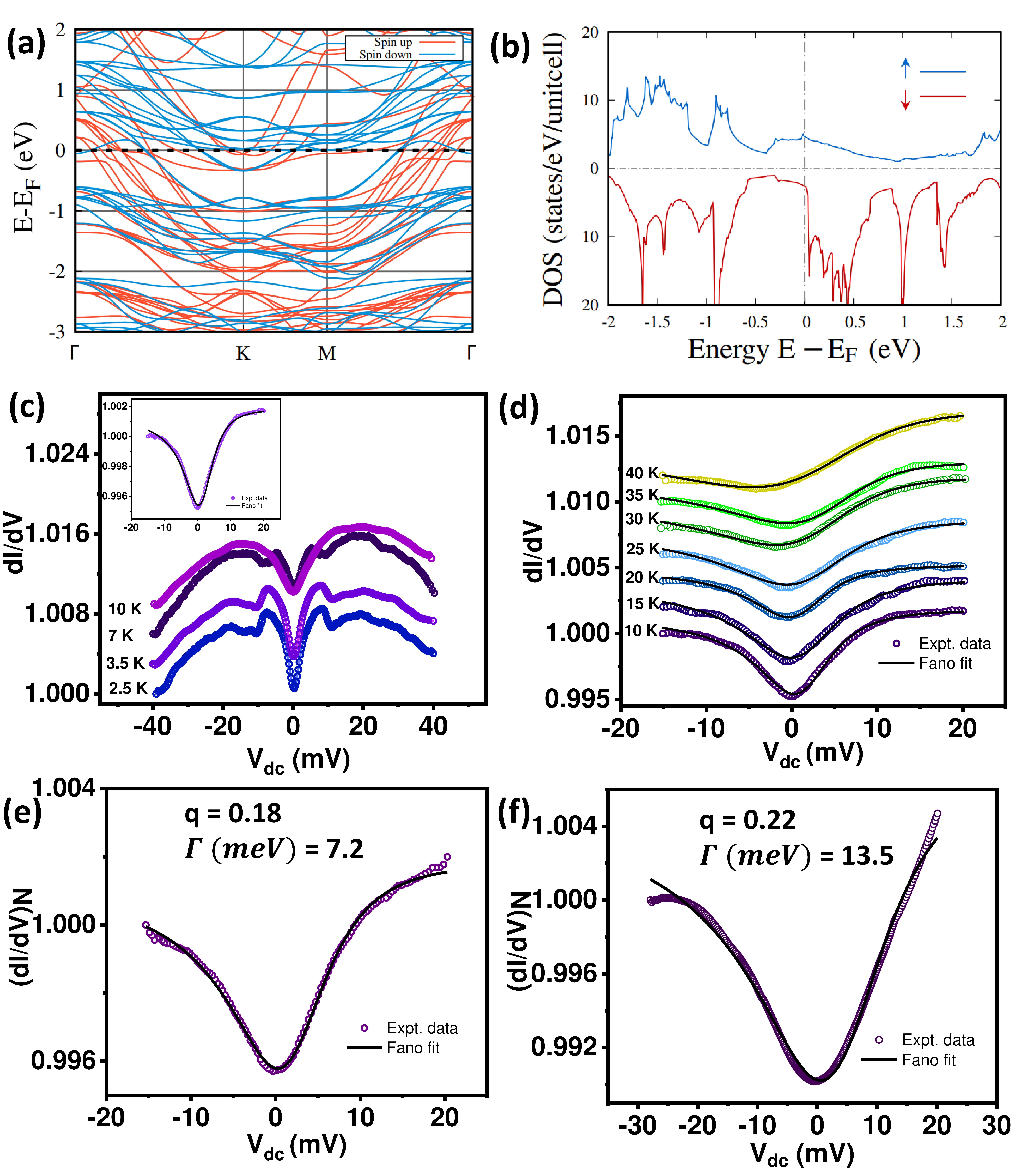}

	\caption{{(a)-(b) Band structure and the electronic density of states (DOS) is calculated for the ferromagnetic state respectively. (c) The temperature dependence of high Z point contact. The inset shows the Fano lineshape fitting (shown by black line) of the conductance spectrum (shown by colored dots) taken at 10K. (d) Temperature dependence of the normal state along with the Fano lineshape fits. Spectra are vertically shifted for clarity. (e)-(f) Three normalised conductance spectra(shown by colored dots) and their Fano lineshape fitting (shown by black line) taken at 12K.
  }}
\end{figure}

The above observation motivated us to investigate the possibility of any other physics competing and/or cooperating with spin-polarized Andreev reflection at the Nb/Fe$_3$GeTe$_2$ interfaces. For that, we gradually increased the temperature of the point contact for a high $Z$ contact, as shown in Figure 3(c), and noted the spectral features over a larger bias range ($\pm40 mV)$. As it was mentioned before, the central dip in an Andreev reflection spectrum is primarily due to non-zero $Z$ and $P_t$, and that should disappear at the $T_c$ of the superconductor forming the junction. However, we found that the central dip structure remained even above 10 K, definitely above the $T_c$ of Nb. Moreover, the normal state spectra also displayed broad conductance peaks around $\pm18 mV$. These peaks compete with the Andreev reflection features in the superconducting state thereby causing anomalous features near 15 mV, where the background gives a downward trend to $dI/dV$ with decreasing $V$, while the peaks due to Andreev reflection give an upward trend. It is this competition that brings the zero-bias conductance of the 10K spectrum below the 7 K spectrum. In this context, we note that the relative strength of the background anomaly and the enhancement due to Andreev reflection are seen to be different at different points. A visual inspection of the spectra presented in Figure 3(c) and Figure 2(a)-(c) clearly reveals this difference.

Now it is important to understand the special features in $dI/dV$ that appeared above the $T_c$ of Nb. A careful inspection of the normal state data reveals an asymmetry between the $\pm V$ regions. As shown in the inset of Figure 3(c), the asymmetry observed in our data at 10 K could be fit well with a Fano line-shape \cite{Fano}. The line-shape was generated using the formula:\begin{equation}
dI/dV \propto \frac{(\epsilon+q)^2}{1+\epsilon^2};  \epsilon = \frac{eV-\epsilon_0}{\gamma}
\end{equation} Here, $V$ is the dc bias, $q$ is the asymmetry factor, $\epsilon_0$ is the position of the resonance in the energy scale, $\gamma$ is the resonance at HWHM (half width at half maximum). Since a Kondo lattice behavior has already been reported in Fe$_3$GeTe$_2$ \cite{Zhang,Zhao}, it is rational to attribute the Fano-like normal state feature with Kondo effect in a Kondo lattice . The normal state of various ballistic contacts was investigated and two such normal state spectra along with Fano fitting parameters are shown in the Figure 3(e)-(f). For a Kondo lattice like Fe$_3$GeTe$_2$, under a point contact geometry, within a two-channel model, the special line shape might be due to the interference between the two current paths, one through the channel of the itinerant electrons and the other one through the quasiparticles that have attained a significantly higher mass due to the interaction between the $d$-electron spins and the spins of the itinerant electrons. In Figure 3(d), we investigated the normal state spectra and their Fano line shape as a function of increasing temperature. The resonance width given by $\gamma$ increases with increasing temperature. This behavior is commonly seen in Kondo-induced Fano line shapes of $dI/dV$ spectra. Such Fano lineshape fitting of experimental point contact spectra was also done for a number of heavy Fermionic superconductors in the past, including CeCoIn$_5$ \cite{Park,Fogelstrom} and URu$_2$Si$_2$ \cite{Park3}.

Therefore, from the analysis of the superconducting and the normal state spectra, it is understood that the Andreev reflection features appear in the presence of a strong background due to the Kondo lattice behavior of Fe$_3$GeTe$_2$. The Kondo-related features interplay with the superconductivity related features below the $T_c$ of Nb. Since the Kondo anomaly at zero-bias contributes to the overall conductance drop at $V$ = 0, a modified BTK fit without incorporating Kondo effect is expected to give an overestimate of $P_t$. In other words, an enhanced $P_t$ needs to be used to take into account the additional suppression of the low-bias Andreev reflection due to the presence of the Kondo lattice background. This explains the apparent discrepancy between the experimentally measured parameters with the estimates using the band structure calculations.

 DR thanks DST INSPIRE for financial support. GS acknowledges financial support from Swarnajayanti Fellowship awarded by the Department of Science and Technology, Govt. of India (grant number: \textbf{DST/SJF/PSA-01/2015-16}).

%%%%%%%%%%%%%%%%%%%%%%%%%%%%%%%%% 

\end{document}